\documentclass[preprint]{JHEP3} 

\usepackage{epsfig,multicol}
\usepackage{amsmath}
\newcommand{\met}{\mbox{$\protect \raisebox{.3ex}{$\not$}E_T$}}
\newcommand\fverb{\setbox\pippobox=\hbox\bgroup\verb}
\newcommand\fverbdo{\egroup\medskip\noindent%
			\fbox{\unhbox\pippobox}\ }
\newcommand\fverbit{\egroup\item[\fbox{\unhbox\pippobox}]}
\newbox\pippobox

\title{Applying Rule Ensembles to the Search for Super-Symmetry at the Large Hadron Collider}

\author{J.~Conrad  \\ The Royal Institute of Technology (KTH), Physics Department, AlbaNova University Center, 10691 Stockholm, Sweden \\ PH-Department, CERN, CH-1211, Geneva 23, Switzerland.}

\author{F.~Tegenfeldt \\ Iowa State University, Ames, IA 50011-3160, USA}

\received{February 20, 2001} 		
\revised{May 1, 2001}
\accepted{November 27, 2001}		

\preprint{\hepph{0605106}}	

\abstract{In this note we give an example application of a recently presented predictive learning method called {\it Rule Ensembles}. The application we present is the search for super-symmetric particles at the Large Hadron Collider. In particular, we consider the problem of separating the background coming from top quark production from the signal of super-symmetric particles.\\
The method is based on an expansion of base learners, each learner being a rule, i.e. a combination of cuts in the variable space describing signal and background. These rules are generated from an ensemble of decision trees. One of the results of the method is a set of rules (cuts) ordered according to their importance, which gives useful tools for diagnosis of the model. We also compare the method to a number of other multivariate methods, in particular Artificial Neural Networks, the likelihood method and the recently presented boosted decision tree method. We find better performance of {\it Rule Ensembles} in all cases. For example for a given significance the amount of data needed to claim SUSY discovery could be reduced by 15 \% using {\it Rule Ensembles} as compared to using a likelihood method}

\begin{document}


\section{Introduction}
Multivariate methods are widely used in high energy physics, for a review, see for example \cite{Prosper:2002hw} and references therein. A typical task is the classification of a data set into two classes, often signal and background. With few exceptions, the concept of a multivariate method can be formulated as follows: The goal is to find the function $F$
for which
\begin{equation}
y = F(\vec{x})
\end{equation}
where $y$ is  a target value and $\vec{x}$ denotes the vector of input values. In particle physics applications with the task of separating signal from background, $y$ encodes signal and background class (for example $+1$ for signal and $-1$ for background), and $\vec{x}$ is a set of observables ( for example the transverse momentum of particles or reconstructed mass) used to distinguish between these two classes. Finding this function is a minimization problem
where
\begin{equation}
\label{eq:min}
F_{opt}(\vec{x}) = \arg{\min_{F}{E_{\vec{x}y}L(y,F(\vec{x}))}}
\end{equation}
Here $E_{\vec{x}y}$ denotes the expectation value over the distribution of all variables in the data to be 
characterized.
$L(y,F(x))$
is the cost function, i.e. the penalty for predicting a value $y$ when a different value $y_{true}$ is the true value. Different multivariate methods differ in the way the cost
function $L$
is defined and how the function $F$ is constructed.
Commonly the approximation to $F$ is found by applying some learning algorithm to a set of known cases, a ``training'' sample. 
In this note we will describe and apply a learning method which is based on the idea of ``ensemble learners'', called {\it Rule Ensembles (RE)}, which has been presented recently by Friedman \cite{Friedman:2005},\cite{Friedman:2005b}. \\

\noindent
The paper is organized as follows: in the next section we will give a brief introduction to the method of ensemble learning machines and the particular implementation of {\it RE} considered here. For more details on this method we refer the reader to \cite{Friedman:2005} and \cite{Friedman:2005b}. In section \ref{sec:application} we will describe the search for super-symmetric particles and in particular rejection of top background. In section \ref{sec:performance} we show how the method performs on the problem considered, followed by a comparison with other common methods in section \ref{sec:comparison}. The final section is devoted to conclusions.

\section{Ensemble learning machines}
\label{sec:ensemblelearning}
Ensemble learning machines are constructed by building linear combinations of learning machines (in that case called ``base learners''). Assuming that the prediction value of the $m^{th}$ base learner is given by $f_m(\vec{x})$ then the prediction value of the ensemble is given by
\begin{equation}
\label{eq:base}
F(\vec{x}) = a_\circ + \sum_{m=1}^M a_m f_m(\vec{x})
\end{equation}
$f_m(\vec{x})$ could be for example the prediction of the $m^{th}$ decision tree in an ensemble of $M$ decision trees. The minimization problem thus becomes the problem of finding the coefficients in this linear expansion in base learners.\\
A recently published example an ensemble learning machine (though the authors did not present in the context of ensemble learners) is the method of boosted decision trees \cite{Roe:2004na} which in that example were shown to perform better than Artificial Neural Nets (ANN). 

\subsection{Rule based ensembles}
\noindent
The base learners presented in this note are of the form:
\begin{equation}
r_m(\vec{x}) = \prod_{j=1}^{n} I(x_j \in s_{jm})
\end{equation}
where I($\cdot$) indicates the truth of its argument and $s_{jm}$ indicates a subset of all possible values of variable $x_j$.
Thus, $r_m (x)$ is either 1 or 0 depending on whether or not $x_j$ falls into its $s_{jm}$. We will call $r_m(x)$ a {\it rule} in the remainder of this paper. A typical rule in context of particles physics could for example be ($n=1$):
\begin{equation}
r_m(N_{el})= I(2 < N_{el} < 5)
\end{equation}
i.e.,
the number of electrons is larger than 2 and smaller than 5. The variables considered could be both numerical (usually the case in physics applications, like the above example) or categorical (for example, job status $\in$ \{tenured, post-doc,unemployed\}).
With rules as base learners
, equation \ref{eq:base} writes
\begin{equation}
\label{eq:base1}
F(\vec{x}) = a_\circ + \sum_{m=1}^M a_m r_m(\vec{x})
\end{equation}
For completeness, we mention that linear functions of the input variables are particularly difficult to approximate by such a set of rules, which is why the particular implementation of {\it RE} presented here includes optionally a term which explicitly includes a dependence on the input variables $\vec{x}$:
\begin{equation}
\label{eq:base2}
F(\vec{x}) = a_\circ + \sum_{m=1}^M a_m r_m(\vec{x}) + \sum_{j=1}^n b_j l_j (x_j)
\end{equation}
where $l_j(x_j) = \min{\left(\delta_j^+,\max(\delta_j^-,x_j))\right)}$. The $\delta_j^{+,-}$ functions are  the $\beta$ and (1-$\beta$) quantiles of the data distribution (taken over the $N$ events for each $x_j$), respectively. These functions assure that in case of outliers instead of the outlier a quantile of the distribution is chosen. Values for $\beta$ are generally small, $\beta \sim$ 0.025.

\subsection{Ensemble generation and Minimization}
\label{sec:ensembles}
The particular algorithm to generate the ensemble and to find the coefficients can differ. In the case presented in this paper, the algorithm used is based on the importance sampled learning ensemble (ISLE) methodology described in \cite{Friedman:2003}. Here we restrict ourselves to give a short summary only. For a given set of base learners $\{r_m(\vec{x})\}^M_1$ a linear regularized regression is performed on the training data. 
\begin{equation}
\{\hat{a}\}_0^M = \arg{\min_{\{a_m\}^M_0}} \sum_{i=1}^N{L\left(y_i,a_0 + \sum_{m=1}^M a_m r_m(\vec{x}_i)\right) + \lambda \sum_{m=1}^M|a_m| }
\label{eq:coefficients}
\end{equation}
where we omit the linear term for simplicity. The first term in equation \ref{eq:coefficients}
is the previously
mentioned measure of the prediction risk (see equation \ref{eq:min}) and the second term is the ``regularization'' term which penalizes large values of the coefficients for each base learner.
The ensemble of rules is generated using a decision tree. Each decision tree is viewed as a collection of rules and used as a base learner $r_m(\vec{x})$. In order to understand a decision tree it is best to describe how it can be constructed. The first step is to select one of the input
variables and loop over all events. For each
event value of the discriminant variable divide the training sample into two, one containing all events with a value less than the
considered value and one with
all events having a value larger than the considered value. The event value which gives best separation between signal and background is chosen and the procedure is repeated for all discriminant variables considered. The discriminant variable and its event value giving overall best separation is used to form the first decision in the decision tree. The training sample is thus divided into two new samples ({\it branches}) corresponding to the ones produced by this decision.
This procedure
is then repeated for each branch in turn, recursively until a predetermined number of final branches is obtained
or until the branch consists of pure signal or background.
For a more detailed description of decision trees we refer the reader to for example \cite{Breiman:1985}.\\
From equation \ref{eq:coefficients} it can be seen that there is the need for defining a cost function. The cost function chosen in the example presented in this note is:
\begin{equation}
L(y,F) = [y - \min(-1,max(1,F))]^2 
\end{equation}
which has been shown to perform comparable to other commonly used methods (e.g. \cite{Friedman:2004}).\\
Calculations presented in this note have been performed within the $R$ statistics framework \cite{R} using the RuleFit package presented in \cite{Friedman:2005}.
For the comparison between multivariate methods, the software package TMVA~\cite{TMVA} was used.

\section{Search for super-symmetric particles in ATLAS}
\label{sec:application}
\subsection{Introduction}
Super-symmetry (SUSY)~\cite{SUSY} is an extension of the Standard 
Model. It
is particularly attractive since it provides a solution to the the so-called hierarchy problem and also offers the possibility to incorporate gravity into a quantum field theory.  
SUSY postulates the existence of superpartners to all known particles. The superpartners of bosons are fermionic (spin $n/2$, where $n$ is an integer ) whereas the fermion partners are bosonic (spin $n$). 
SUSY particles have a quantum number called R-parity. If this quantum number is conserved, there is consequently a stable SUSY particle which will not decay further, the lightest stable particle (LSP). Experimental studies of SUSY are complicated by the fact that a large amount of free parameters are added to the SM, the minimal extension having 106 free parameters. SUSY is therefore often (and also in this note) studied in a limit which is simplified at the Planck scale (so-called SUGRA) ~\cite{SUGRA,PDG}. In this approximation, the number of extra free parameters to be considered reduces to 5. The remaining parameters are the universal scalar masses $m_0$, the gaugino masses $m_{1/2}$, a trilinear coupling $A_0$, the ratio $\tan \beta$ and a sign sgn$\mu$.


\noindent
The mass scale of this theory is dependent on the electroweak breaking scale which is given by the Higgs mass.
Currently, the experimental lower limit of the Higgs mass is 114.4 $GeV/c^2$~\cite{PDG} and from theoretical arguments, its upper limit is about 800 $GeV/c^2$.
Consequently, SUSY could be discovered at the TeV scale, thus being quickly accessible at the Large Hadron Collider (LHC) ~\cite{Bruning:2004ej}.
The LHC is a proton-proton collider currently being built at the European Center for Nuclear Research (CERN) outside Geneva. The envisaged center of mass energy in proton-proton collisions is 14 TeV and the design luminosity will be  $\mathcal{L} = 10^{-2} pb^{-1} s^{-1}$. At the LHC, there will be  4 multi-purpose experiments: ATLAS~\cite{Armstrong:1994it}, CMS~\cite{unknown:1994pu}, LHCb~\cite{Amato:1998xt} and ALICE\cite{:1995pv}. ATLAS and CMS focus on the discovery and precision measurements of the Higgs sector and SUSY. LHCb is designed for B-physics. Part of the LHC program includes collisions of heavy ions (lead), which are the main objective of ALICE. In this paper, we will focus on SUSY studies using the ATLAS detector.

\subsection{The samples}

The ATLAS collaboration considers several points in SUGRA parameter space.
In this paper we will only consider two of them, referred to as SU1 and SU3.
 Their SUGRA parameters are summarized in table~\ref{tab::susyModels}.

\TABLE[htbp]{

\begin{tabular}{c|l|l|l|l|l|l}
\hline\hline
Model & $\sigma_{LO}$ (pb)& $m_0$ ($GeV/c^2$) & $m_{1/2}$ ($GeV/c^2$) & $A_0$ & $\tan \beta$ & $sgn \mu$\\\hline
SU1   & 6.8		  & 70		      & 350		      & 0     & 10	     & + \\
SU3   & 19.3		  & 100		      & 300		      & -300  & 6	     & + \\\hline
\end{tabular}
\caption{SUGRA parameters and leading order cross-sections for models used in this paper.}
\label{tab::susyModels}
}
The detector response for each sample has been obtained through a full detector simulation using GEANT4 ~\cite{GEANT4}.\\

\subsubsection{Signal}

Irrespective of SUSY model, signatures are complex multijet events possibly involving hard leptons and heavy quark jets. In the models considered here, R-parity is conserved and therefore each SUSY event produces stable heavy and weakly interacting particles that escape undetected. One of the main characteristics for these types of event is therefore large missing transverse energy, \met.


\subsubsection{Background}
Several background sources are currently considered by the ATLAS collaboration. An overview of the main sources is given in table~\ref{tab::bkgSummary}.

\TABLE[htbp]{
\begin{tabular}{c|l|l}
\hline \hline
Source	& $\sigma$ (pb)	& Characteristics \\\hline
QCD	&$1.5\cdot 10^9$& Small \met\ but very large cross-section \\
Top	& 580		& Multijet and large \met\ (escaping hard neutrinos) events \\
W+jets	& 1200		& Similar to top but \met\ less hard \\
\hline
\end{tabular}
\caption{Main background sources to SUSY. Although top has the smallest cross section, it is the hardness of \met\ (and therefore more SUSY like) which makes top likely to be the largest background.}
\label{tab::bkgSummary}
}

The relevant decay chain in top events is $t \rightarrow bW$ where the $W$ can either decay hadronically or leptonically. Leptonically decaying $W$ is
particularly difficult to distinguish from SUSY since it is characterized by large \met\ from escaping neutrinos. This is the topology considered in this paper.


\subsubsection{Discriminators}
In this study we considered fourteen discrimination variables. Except for the previously mentioned \met, we considered various combinations of jet momenta as well as the b-jet content. Top events will predominately have 2 b-jets whereas in SUSY several topologies for 0,1,2 or more b-jets are possible. A summary of the discriminators used is given in table~\ref{tab:discriminators}.\\

\TABLE[htbp]{
\begin{tabular}{c|l}
\hline\hline
Discriminator	& Description \\\hline
\met		& Missing $E_T$ \\
$M_{\mathit{eff}}$ & \met$+\sum_{jet} p_T^{jet}$ \\
$p_T^{0}$, $p_T^{1}$ & First two hardest jets \\
$H_T$		& $\sum_{jet} p_T^{jet}$ \\
$H_T^{3j}$	& $H_T - p_T^0 - p_T^1$ \\
$H_E$		& $\sum_{jet} E_{jet}$ \\
(A)planarity, sphericity & Combinations of eigenvalues of jet momentum tensor \\
$N_{jets}$	& Number of jets \\
$N_{bjets}$	& Number of b-jets \\
$N_e$, $N_{\mu}$& Number of electrons/muons \\
\hline
\end{tabular}
\caption{The discriminating variables used in the separation of SUSY signal from top background}
\label{tab:discriminators}
}

\FIGURE[c]{
\epsfig{file=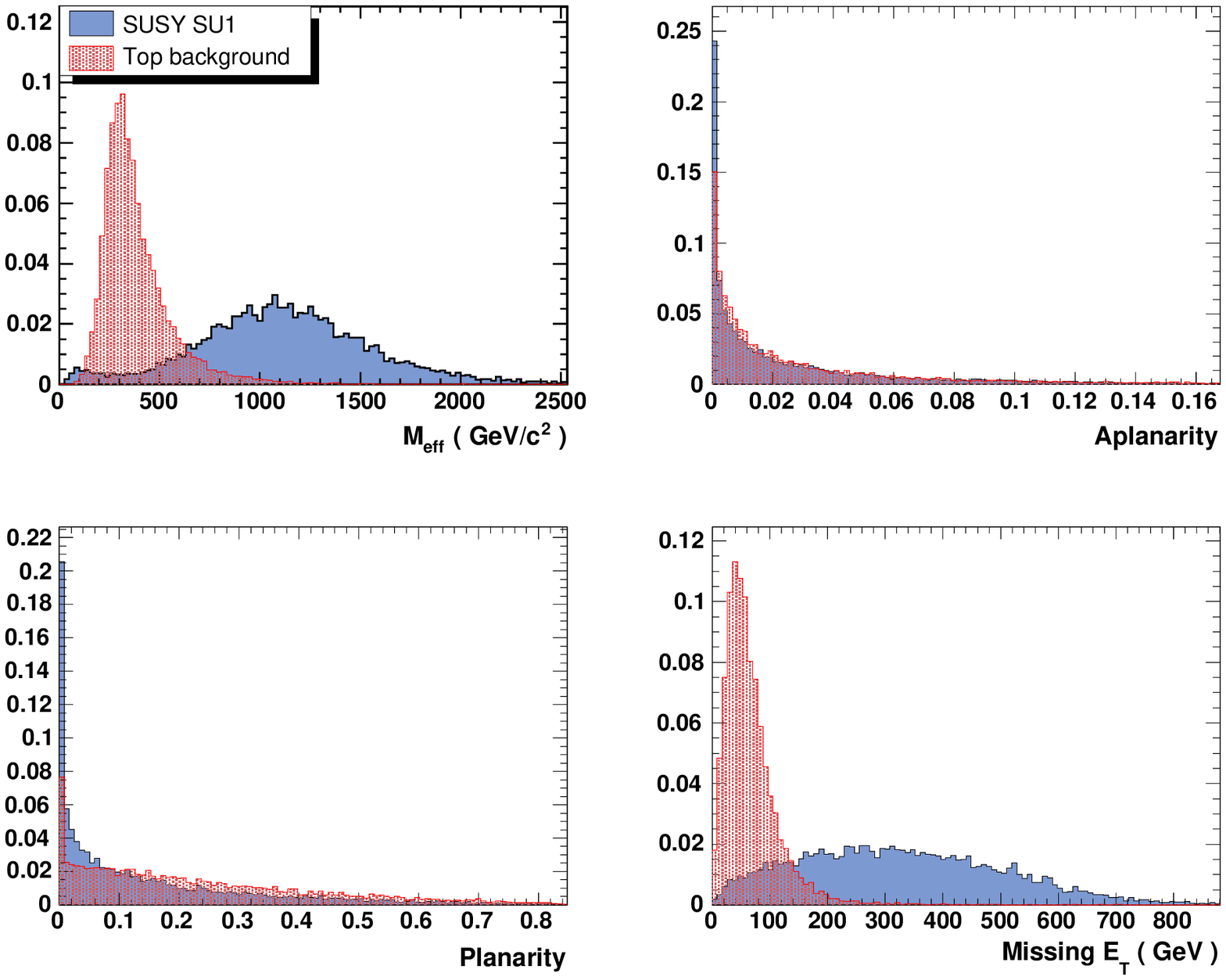,,height=10cm,width=12cm} 
\caption{Some example plots of discriminators used in the separation of SUSY signal from top background}%
\label{fig:discriminators}
}


\noindent
The difficulties constraining SUSY models results constitutes an unavoidable systematic effect: the SUSY model used to train the learning machine might not be the one nature has chosen. Training and testing on two different SUSY models thus yields information about what the effect of training on the wrong model might be as well as give us information about how stable {\it RE} are under systematic shifts of the input variables. The difference in one of the discrimination variables, $M_{\textit{eff}}$ (defined in table \ref{tab:discriminators}) can be seen in figure \ref{fig:Meff}. The distribution is shifted by about 20 \%.

\begin{center}
\FIGURE[c]{
\epsfig{file=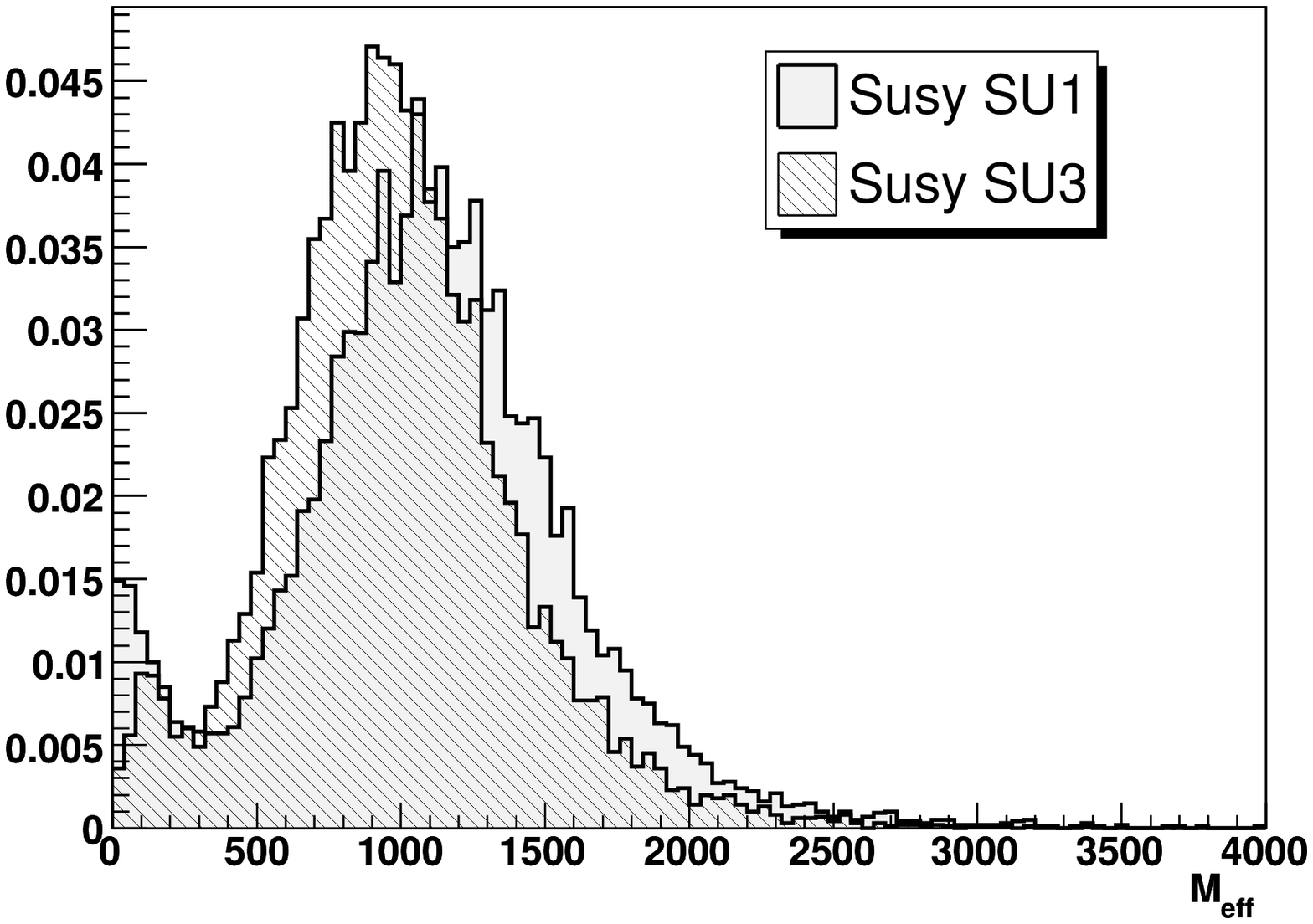,height=10cm,width=12cm} 
\caption{The discriminator  $M_{\textit{eff}}$ (defined in table \ref{tab:discriminators}) for the SU1 and SU3 benchmark models}%
\label{fig:Meff}
}
\end{center}

\section{Performance of the method}
\label{sec:performance}
The performance of the method can be measured in different ways. We will here present two ways of assessing the performance. Firstly, cross validation is used to estimate the absolute error rate, average error rate and area under the ROC curve. The ROC curve is just the correlation between negative error rate (i.e. class signal is misclassified as background)  and positive error rate (i.e. class background is misclassified as signal). Cross validation is a common method in which the training sample is randomly subdivided into sub-samples on which the model is mutually trained and tested. Table \ref{tab:xval} compares the results for the model which was trained on  SU1 SUSY model only
and one which was trained on a mix between SU1 and SU3. Also included are the results for a model where we only use the four most important variables of the original sample (see section \ref{sec:varimp}).\\

\noindent
From this table, it can be seen that the rate of top events misclassified as SUSY events is slightly higher than vice versa. It can also be seen that the method only becomes slightly worse. For example the average error rate increases by 0.3 $\%$ in case only the four most important variables are used. Worst performance of the method (though also only slighlty worse) is obtained if the signal sample is in itself consisting of two distinct classes (in this case SU1 and SU3).\\

\noindent
Of more practical interest for particle physicists is the performance of the method on independent test samples. In particular score distributions give an intuitive impression on how good the methods perform.  We show the obtained scores for the model trained on SU1 and tested on independent test samples consisting of SU1 events, SU3 events and top events in figure \ref{fig:scoresall}.

\TABLE[htbp]{
\begin{tabular}{l|l|l|l|l}
\hline\hline
Model 	            &1-$A_{ROC}$ & neg. error rate & pos. error rate & aver. error rate   \\ \hline
SU1 (14 variables)  &0.024       &0.064            & 0.082           & 0.073              \\
MIX (14 variables)  &0.029       &0.072            & 0.092           & 0.082              \\
SU1 (4 variables)   &0.026       &0.068            & 0.085           & 0.076              \\ \hline
\end{tabular}
\caption{Error rates and area under the ROC curve for the samples considered in this study}
\label{tab:xval}
}

\noindent
These score distributions can be used to calculate efficiency for signal and background as a function of cut value in the score. In figure \ref{fig:eff} we show the correlation between background and signal efficiency for the combinations (trained on SU1, tested on SU1 and SU3 as well as trained on the mixed sample and tested on SU1 and SU3).\\

\noindent
Best performance is non-unexpectedly found in the case that we train on SU1 and test on SU1 (SU1-SU1). Worst performance is found if we train on a mixture of SU1 and SU3 and test on SU3. The effect of a systematic shift in signal modeling can be seen from the difference of the curve where we trained on SU1 but tested on SU3, which gives the next to worst result. For example at a signal efficiency of 80 \% the background contamination increases by roughly 50 \% (from $\sim$ 2 \% to $\sim$ 3 \%), still a relatively moderate increase. An interesting case is the case where we train on a mixture of SU1 and SU3 but test on SU1. Here, for large signals efficiency behaves equally well as the case (SU1-SU1), whereas it approaches the case where we train on SU1 and test on SU3 for small signal efficiencies.

\begin{center}
\FIGURE[c]{
\epsfig{file=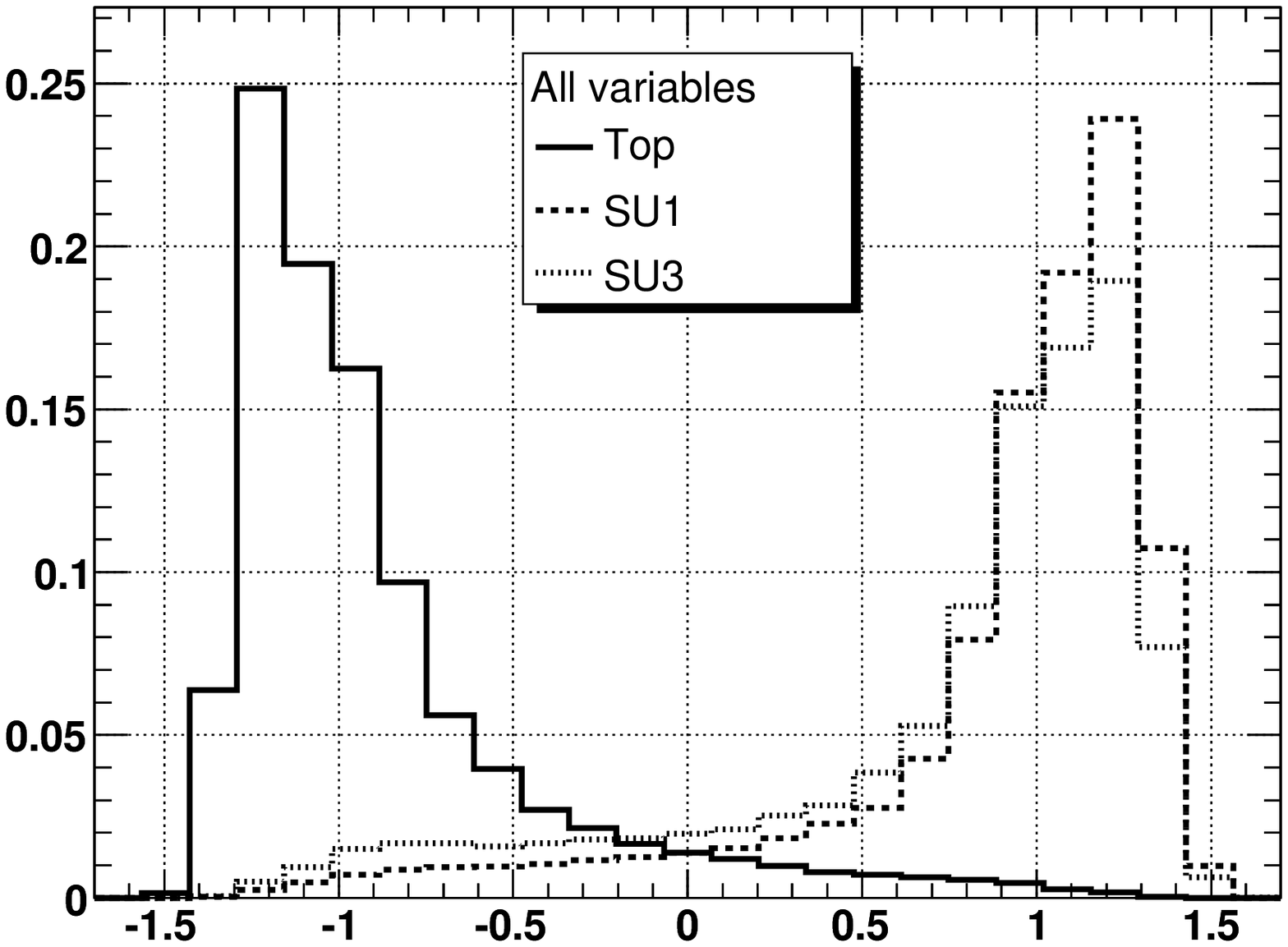,height=10cm,width=12cm} 
\caption{Scores for the case training on SU1 sample and testing on SU1, SU3 and top sample.}%
\label{fig:scoresall}
}
\end{center}


\begin{center}
\FIGURE[c]{
\epsfig{file=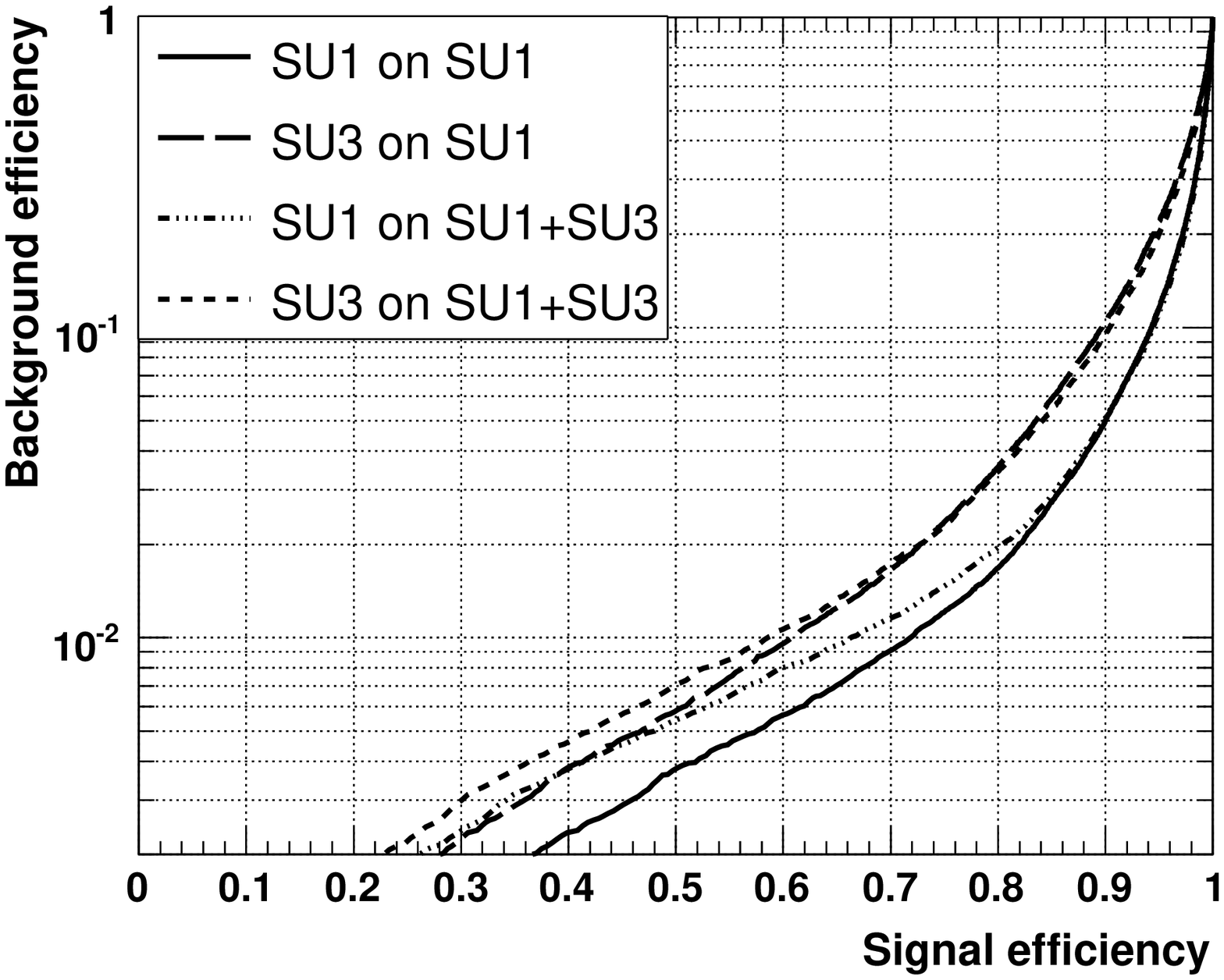,height=10cm,width=12cm} 
\caption{Background vs. signal efficiency for the different samples, trained on SU1 and an equal mixture of SU1+SU3 and tested on SU1 and SU3, respectively.}%
\label{fig:eff}
}
\end{center}


\subsection{Diagnostics}
In this section we illustrate how diagnostics of this method can be used to arrive at a simpler, still almost as efficient characterization of the data.
\subsubsection{Rule importance}
The importance of the rules is defined by the absolute value of their respective coefficients, $|\hat{a}_k|$ and their support, $s_k$. For the rules this is expressed as
\begin{equation}
I_k = |\hat{a}_k \cdot \sqrt{s_k(1-s_k)}|
\end{equation}
The support for each rule is defined as its average response of for the given data.
\begin{equation}
s_k = \frac{1}{N} \sum_{i=1}^{N} r_k(\vec{x}_i)
\end{equation}
For the linear predictors the importance becomes
\begin{equation}
I_j = |b_j| \cdot std(l_j(x_j))
\end{equation}
where $std()$ is the standard deviation over data.
We summarize the four most important rules in table \ref{tab:rule_importance}.\\
\TABLE[htbp]{
\begin{tabular}{ll|l|l}
\hline\hline
Rule 	& &Coefficient &Relative Importance   \\ \hline
1  &246.8 $< \met < $2180 &0.1539 &100.0 \\ \hline
2  &20.5 $ <p_{T,2}^{jet} < $ 236.2&-0.1371 &96.1 \\
   &$\met < $ 145.5 & & \\ \hline
3  &142.5 $<M_\mathit{eff}<$ 738.6 &-0.1369 & 95.8 \\
   &$\met < $ 205.3 & & \\ \hline
4  &139.4 $<M_\mathit{eff}<$ 681.5 &-0.1215 & 85.2 \\
   &$\met < $ 193.5 & & \\ \hline
\end{tabular}
\caption{The four most important rules using all variables trained on SU1 signal.}
\label{tab:rule_importance}
}

\noindent
The cut in $\met$ is (not unexpected) the most important cut, followed by a combination of $\met$ and the $p_{T}$ of the next-to hardest jet. While in the case of SUSY search this is a almost trivial statement, in more complicated cases this might be valuable information by giving hints on which variables should be paid special attention to and in which region of the parameter space. For example, one could find out that the cuts are performed in a region of the parameter space where the quality of the variable (say for example, resolution in $p_{T}$) is expected to be worse than in other regions. An improvement of the quality of the variable in that region could then be attempted.

\subsubsection{Input variable importance}
\label{sec:varimp}
Another important diagnostic of the model is the relative importance of the input variables to the model. This figure is connected to the rule importance since variables which participate in the most important rules will be judged more important than those participating in less important rules. The importance will also be determined by the frequency in which the variables participate in the rules. The corresponding measure is written as
\begin{equation}
J_l = I_l + \sum_{x_l \in r_k} I_k/m_k
\end{equation}
where $m_k$ is the number of variables participating in rule $r_k$.
The ten most important input variables are shown in figure \ref{fig:varimp}.
\FIGURE[c]{
\epsfig{file=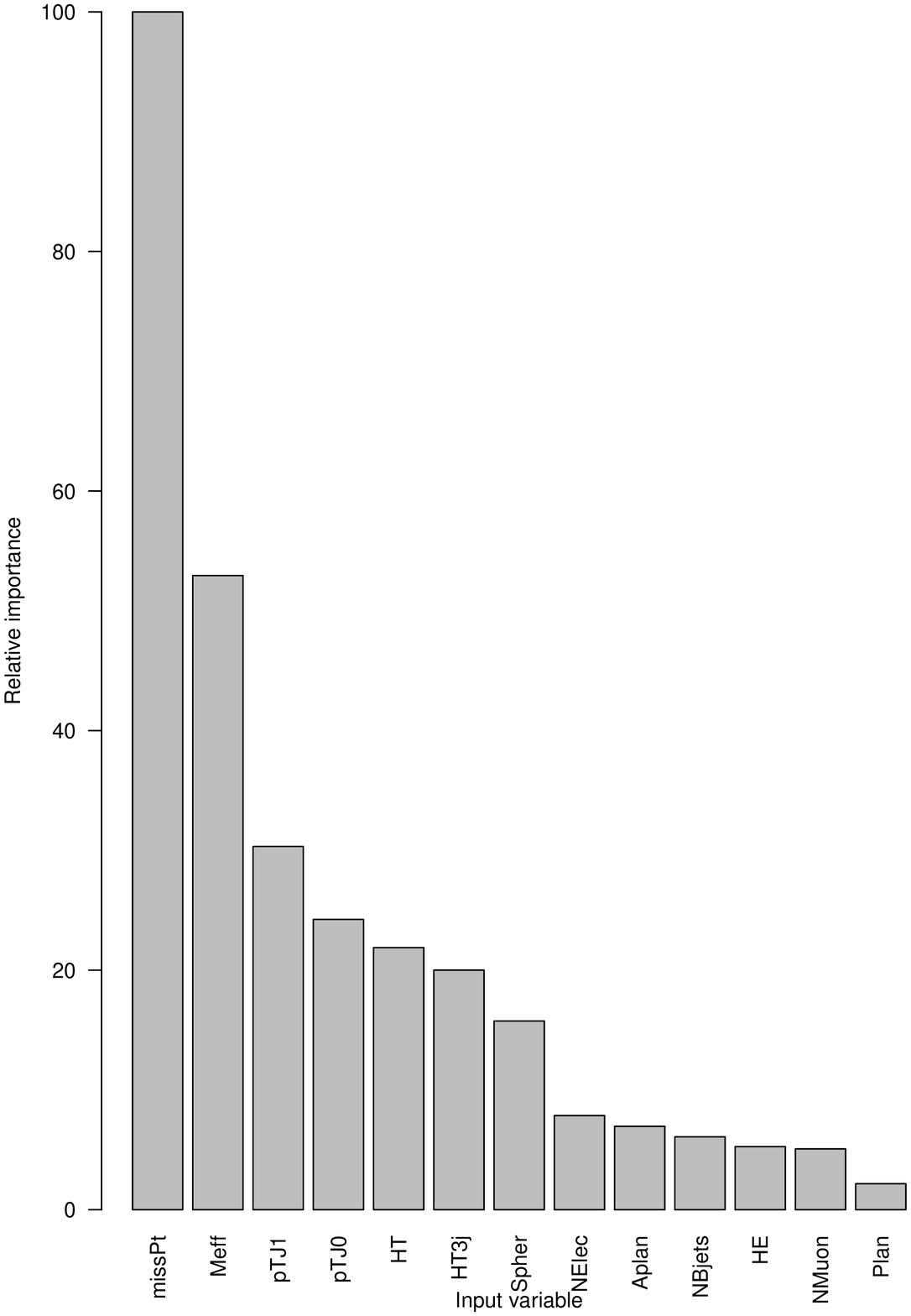,height=10cm,width=12cm} 
\caption{The importance of the variables participating in the sample}%
\label{fig:varimp}
}
 From this figure it is clear that the most important variables are $\met$, $M_\mathit{eff}$, and $p_T$ of the hardest and the next to hardest jets.\\

\noindent
One possibility then arises by constructing a much simpler model only including these most important variables. Indeed, the area of the ROC curve becomes only negligibly smaller (see table \ref{tab:xval}). The corresponding efficiency curves calculated from testing on independent test samples are compared in figure \ref{fig:eff4only}.\\

\noindent
From figure \ref{fig:eff4only} it can be concluded that the model including only the four most important variables is performing only marginally worse than the model including all fourteen discriminating variables. In particular, the degradation in performance is negligible compared to the degradation which is introduced by training on the ``wrong'' super-symmetric model.


\FIGURE[c]{
\epsfig{file=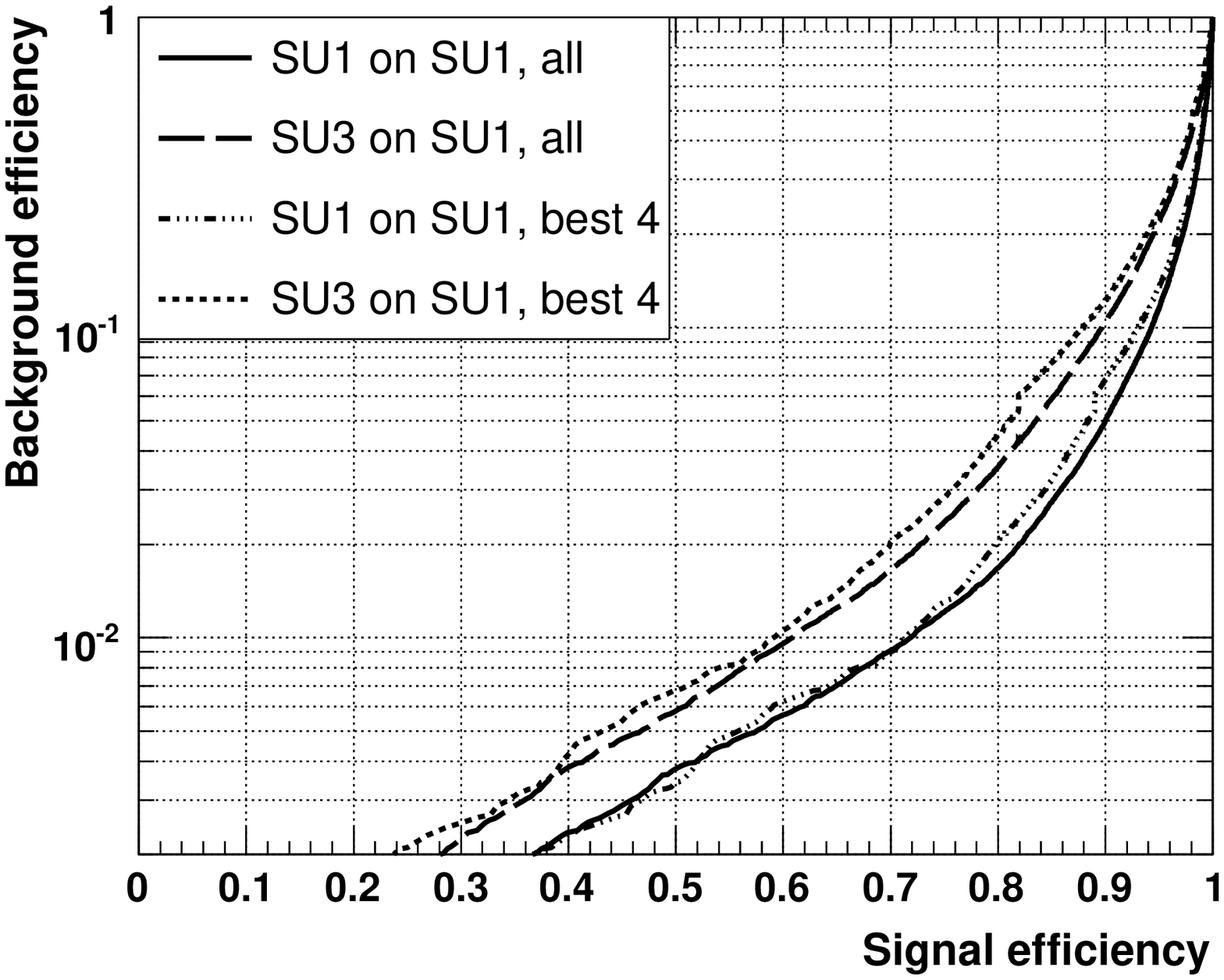,height=10cm,width=12cm} 
\caption{Background vs. signal efficiency for the model using only the four most important variables as compared to the model using all variables}%
\label{fig:eff4only}
}

\section{Comparison with other methods}
\label{sec:comparison}
In general, it is difficult to compare different multi-variate methods, since it is not simple to  prove that in all cases optimal parameters where used. We therefore assume the ``out of the box'' approach, i.e. we compare different methods using recommended default values. For {\it RE} all results presented so far are essentially with default parameters of the machine, i.e. with a minimal amount of tuning. In this section we use the package TMVA~\cite{TMVA} to compare {\it RE} with other common methods. In particular, we consider the Likelihood method, an Artificial Neural Network, the H-Matrix method, Boosted Decision Trees and the Fisher Discriminant.  The following subsections are devoted to a brief description of each method. The vector $\vec{x} = \{ x_0, x_1, ..., x_{N-1}\}$ denotes the $N$ different discriminating variables used. $F(\vec{x})$ denotes, as before, the test statistic used for the separation between the classes. The final subsection will describe the results of the comparison.

\subsection{Description of the different methods}
\subsubsection{Likelihood}
For each event we can define a likelihood for it to be a signal and a background event
through the product of the probability densities of each discriminating variable
\begin{equation}
\mathcal{L}(\vec{x}) = \prod p_i(x_i)
\end{equation}
The probability density functions are determined using a training sample.
As a test statistic, the likelihood ratio is used
\begin{equation}
F(\vec{x}) = \frac{\mathcal{L}_S(\vec{x})}{\mathcal{L}_S(\vec{x}) + \mathcal{L}_B(\vec{x})}
\end{equation}
where the subscripts S and B denotes signal and background class, respectively.
For this method it is assumed that all discriminators are uncorrelated (i.e. independent given the class label).
Linear correlations can be included by first diagonalizing the correlation matrix.
The likelihoods can then be constructed in the transformed variable space where the
correlations are minimal.

In the comparison presented below, the method assuming uncorrelated discriminators is used.

\subsubsection{H-matrix}
In this method, $\chi^2$ estimators are obtained for the signal and background samples
\begin{equation}
\chi^2(\vec{x}) = (\vec{x} - \vec{\overline{x}})^T C^{-1} (\vec{x} - \vec{\overline{x}})
\end{equation}
The test statistics is then constructed from the ratio
\begin{equation}
F(\vec{x}) = \frac{\chi^2_S(\vec{x}) - \chi^2_B(\vec{x})}{\chi^2_S(\vec{x}) + \chi^2_B(\vec{x})}
\end{equation}

\subsubsection{Fisher Discriminant}
The general form of a Fisher test statistic is a linear combination of the discriminating variables
\begin{equation}
F(\vec{x}) = \sum_i a_i x_i
\end{equation}
where the coefficients $\{a_i\}$ are selected such that $F(\vec{x})$ yields maximum separation between the signal and background classes.
In order to determine $a_i$, the average of each variable and the variable covariance matrix
for the signal class and ditto for background are used. 
In this paper we use the so called Mahalanobis-Fisher discriminant where the covariance matrix between the classes is included.
The coefficient $a_i$ is a measure of the discriminating power of the corresponding variable.

\subsubsection{Artificial Neural Network}

The Fisher discriminant works well for separated classes. If the discriminators are not linear, a more general parameterizations of the test statistic can be considered.
In particular, for the case of a so called multi-layer perceptron, the test statistic is then formed by a linear combination of \textit{nodes} organized in \textit{layers}. In the input layer, the nodes are given by the input variables. The output layer consists of one or more nodes corresponding to the various output classes. In between these layers, an arbitrary number of \textit{hidden} layers with an arbitrary number of nodes can be used.
Each node in a given layer is a function of all nodes in the previous layer.
The test statistic takes the form
\begin{equation}
F(\vec{x}) = s\left(  a_0 + \sum_{i=1}^{n} a_i h_i(\vec{x}) \right)
\end{equation}
where $s(\cdot)$ is the \textit{activation function}, often set to $(1+e^{-x})^{-1}$. The functions $h_i(\vec{x})$ are of the same form as $F(\vec{x})$ but act on the previous layer.
Since a neural network can contain an arbitrary number of layers and nodes, it is possible to make more or less good designs. The chosen design in this paper is a network with 4 input variables and 2 hidden layers with 4 nodes each.

\subsubsection{Boosted Decision Trees}
Boosted decision trees are in detail described elsewhere \cite{Roe:2004na}, here we will constrain ourselves to a short description.
Boosted decision trees is another example of an ensemble learner.
 The base learner is in this case a decision tree and boosting is
the algorithm used to construct the ensemble. We gave a description of decision trees in section \ref{sec:ensembles}. The main idea of the {\it boosting} algorithm consists in a re-weighting of misclassified events. Consider a decision tree constructed in the way described above. Training events which are classified as the opposite class will get an increased ({\it boosted}) weight and then a new tree is built using this new weight. The procedure is repeated many times and in this way an ensemble of decision trees is built. The final score of the event will be determined by following each event through each tree in turn and build an (possibly weighted) average over all the trees in the ensemble. There are a number of different algorithms used to define the new weights of the misclassified events. The particular algorithm that we compare with is called {\it AdaBoost}. We refer the reader to \cite{Freund:1996} or to \cite{Roe:2004na} for more details on this algorithm.

\subsection{Results of the comparison}
In table \ref{tab:methodscomparison} we show the signal efficiency of the different methods at three different background efficiencies as well as the separation, defined as:
\begin{equation}
\frac{1}{2}\int_{-\infty}^{\infty}\frac{(S(x) - B(x))^2}{S(x) + B(x)}dx
\end{equation}
where $x$ is the obtained scores, and $S(x)$,$B(x)$ are the signal and background distributions of the scores, respectively. We also show the significance, defined as
\begin{equation}
\frac{(\bar{x}_S - \bar{x}_B)}{(rms_S)^2+(rms_B)^2} 
\end{equation}              
where $\bar{x}_{S,B}$ and $\bar{x}_B$ are the mean score values found by MVA for signal and background, respectively and $rms_{S,B}$ are the corresponding root mean square values. In figure \ref{fig:effCmp} the signal efficiency as a function of background efficiency is displayed for the different methods. Here only the four most important discriminating variables (as defined by {\it RE}) were used.

\FIGURE[c]{
\epsfig{file=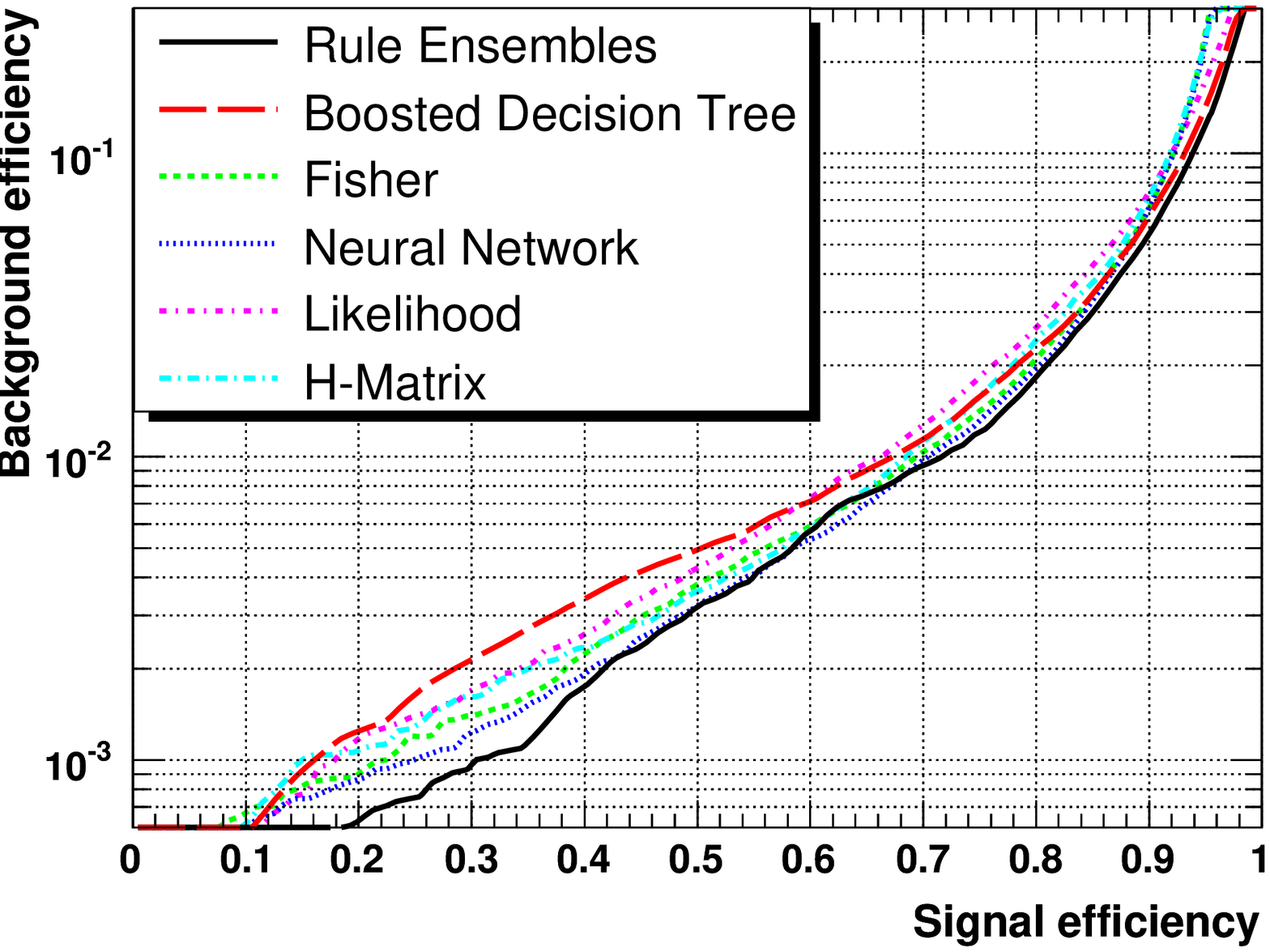,,height=10cm,width=12cm} 
\caption{Comparison of efficiencies between various multi variate methods. Only the four best variables where used.}%
\label{fig:effCmp}
}

\TABLE[htbp]{
\begin{tabular}{l|lll|l|l}
\hline\hline
Method        & \centering{sig. eff.} & &  & significance & separation \\
              & bg = 1 \% & bg = 10 \% & bg = 30 \% & &           \\ \hline
{\it RE.}  & 0.716    & 0.938    & 0.982    & 2.595    & 0.772 \\
Boosted DT    & 0.673    & 0.931    & 0.978    & 2.436    & 0.752 \\
Fisher	      & 0.694    & 0.921    & 0.954    & 1.599    & 0.668 \\
Neural Network& 0.704    & 0.921    & 0.956    & 2.367    & 0.746 \\
Likelihood    & 0.662    & 0.920    & 0.974    & 2.067    & 0.721 \\
HMatrix	      & 0.685    & 0.918    & 0.959    & 2.015    & 0.734 \\
\hline  
\end{tabular}
\caption{Comparison of various multivariate methods. For a definition of the different figures of merit we refer to the text.}
\label{tab:methodscomparison}
}

\noindent
From table \ref{tab:methodscomparison} and figure \ref{fig:effCmp} it can be concluded that {\it RE} performs best for all signal efficiencies and background efficiencies. Most pronounced is the difference, if one requires low background efficiencies. For high signal efficiencies boosted decision trees and ANN perform almost as good as {\it RE}, whereas their performance degrades significantly for low background efficiencies. Overall, ANN come closest in performance to {\it RE}.\\ 

\noindent
In this comparison we choose a sample using only the four most relevant variables. It should be noted, that ANN (and probably also the other methods) in general show degraded performance if irrelevant discriminating variables are added to the sample.

\section{Discussion and Conclusions}
\label{sec:dis}
In this note we illustrate the use of a novel, powerful multi-variate classification method, called {\it RE} to reject top background in search for super-symmetric particles in ATLAS. The method is based on the concept of ensembles of learning machines. In particular, an ensemble of linear combinations of cuts in discriminant variable space is constructed by means of decision trees. Since linear combinations of cuts are used, {\it RE} are easy to diagnose: important cuts and discriminant variables are easily identified, which can be used to simplify the model or to decide which variable deserves extra attention. In our illustration we reduce the initial model of fourteen discriminant variables to a model only including the four most important without a major degradation of the performance of the method.\\

\noindent
Since super-symmetry is so poorly constraint, training samples have to be constructed for benchmark points in the space of possible parameters. We consider in this paper two benchmark models within SUGRA. It is shown that the method is robust under training on the ``wrong'' signal model, indicating stability under systematic shifts of the discriminating variables. Training on the ``wrong'' model degrades the performance moderately. Using a mixture of different benchmark models for training is shown to be not efficient.\\

\noindent
The top background is most probably the dominant background in the search for SUSY since its signature is hardest to discriminate from SUSY events. The presented method shows better discrimination power than the most common multi-variate methods used in high energy physics. The importance of choosing such a method should not be underestimated: the ratio of the amount of data which needs to be taken for a given significance scales like the square of the ratio of signal efficiencies for the two methods. In our example, the amount of data taken using {\it RE} for an expected background efficiency of 1 \% is only 85 \% of the amount of data which would be necessary if using the likelihood method.


\acknowledgments
We have made use of the ATLAS physics analysis framework and data
samples which are the result of collaboration-wide efforts.
The authors would like to thank Jerome Friedman for discussions.
We would also like to thank the TMVA development team for their help.

\end{document}